\documentstyle[preprint,aps]{revtex}

\begin{document}

\draft
\preprint{
  \parbox{2in}{Fermilab--Pub--94/406-T \\[-0.12in] hep-ph/9412250}
}

\title{Finite Width Effects in Top Quark Decays}
\author{Gregory Mahlon \cite{GDMemail} and
        Stephen Parke \cite{SJPemail}}
\address{Fermi National Accelerator Laboratory \\
P.O. Box 500 \\
Batavia, IL  60510 }
\date{December 1994}
\maketitle
\begin{abstract}
Motivated by evidence that the top quark mass lies near the
$b\thinspace WZ$ threshold, we compute the decay rate for
$t\rightarrow b\thinspace WZ$ in the Standard Model,
including the effects of the finite widths of the $W$ and $Z$ bosons.
In the limit where the width effects are negligible,
our results disagree with previously published calculations.
We also examine the decay $t\rightarrow b\thinspace WH$.
Although the widths induce a sizable
enhancement near threshold for both decays, we find that
the rates are still too small to be
observed in the present generation of experiments.
This means that detection of either mode in one
of these experiments would be a signal of new physics.
\end{abstract}


\section{Introduction}

Recent experimental evidence from CDF \cite{CDFtop} as well as the
global fit to LEP data \cite{LEPfit} suggests that the top quark
mass lies near the threshold for decay into a $b\thinspace WZ$
final state.
Decker, Nowakowski, and Pilaftsis (DNP) \cite{Decker}
have presented numerical results for the  decay
$t \rightarrow b\thinspace W^{+}Z$.
The DNP calculation treats the $W$ and $Z$ bosons in the  stable
particle approximation.  In light of the recent
experimental data, it is prudent to reexamine this decay, including
the subsequent decays of the $W$ and $Z$. One expects, and
indeed we find, that there are significant corrections to the
stable particle result for the range of interesting top masses.
Unfortunately, because of strong phase space suppression, the rates
are too small to be seen in the near future, even when the
enhancement from the non-zero $W$ and $Z$ widths is taken into
account.  Thus, should this decay mode of the top be observed,
it would signal physics beyond the Standard Model.

Another process discussed by DNP~\cite{Decker,DeckerH} is
$t \rightarrow b\thinspace W^{+} H$.
We take this opportunity to re-examine this  decay as well,
including the width of the $W$.
As before,
the rate is significantly enhanced near threshold, but not enough
to render the mode visible.


\section{The Decay $\noexpand\lowercase{t} \rightarrow
\noexpand\lowercase{b} WZ$}

We begin with a discussion of the decay chain
\begin{equation}
 t \rightarrow b\thinspace W^{+}\thinspace Z \thinspace ;
\quad W^{+} \rightarrow   \mu^{+} \nu_{\mu}\thinspace ,
\enspace Z \rightarrow   e^{+}e^{-}.
\label{tbWZ}
\end{equation}
There are a total of nine Feynman diagrams contributing to the desired
final state in the Standard Model.  The first three are
the usual diagrams for
$t\rightarrow b\thinspace W^{+}Z$ where the $W^{+}$ and
$Z$ decay to the indicated final states.  The next two diagrams
involve radiation of the $Z$ from one of the decay products of the
$W^{+}$.  These diagrams are expected to be small in comparison
to the first three, as all of the $W$ propagators they contain are
far from mass shell.  The final four diagrams involve replacing the
$Z$ by a photon.  To reduce the contamination from the photon,
we impose a cut on the invariant mass of the $e^{+}e^{-}$ pair.
This is a reasonable approach because
experimentally such a cut would also be implemented.
If we choose a cut of $0.8\thinspace m_Z$,
we find that the contribution
from the photon diagrams is typically less than 2\%.
%
%
Nevertheless, the numerical results we present below include all nine
contributions.

Because the $t$ mass is likely to lie near the threshold for
$b\thinspace W^{+}Z$ production, it is important to include the finite
widths of the $W^{+}$ and $Z$ in the calculation.
We take the propagator of a $W$-boson of momentum $W$, mass $m_W$,
and width $\Gamma_W$  in the unitary gauge to be
\begin{equation}
{
{-i}
\over
{W^2-\hat m_W^2}
}
\biggl(g^{\mu\nu} -
{
{{W^\mu}W^{\nu}}
\over
{\hat m_W^2}
}\biggr)
\end{equation}
where $\hat m_W \equiv m_W + {1\over2}i\Gamma_W$.  We use a similar
expression for the $Z$ propagator.  Note the presence of the
complex mass in the longitudinal part of the propagator:  it is
required to preserve gauge invariance in the process
$t \rightarrow b\thinspace W^{+}\gamma$.
Similarly, it is inconsistent
to put a complex mass in the $t$-quark propagator
unless the production process is included.
We have found that a 1 GeV shift in $m_t$ produces
a shift of 10\%--15\% in the rate for top masses just above the
$b\thinspace W^{+}Z$ threshold.
We have convolved our curve with a Breit-Wigner distribution in $m_t$
to simulate what one might find were production properly included, and
see a change of only a few percent.
Hence, we exclude the
width of the $t$ from our discussion.

In the case where both the $W$ and $Z$ decay to the same lepton
family, there are additional diagrams resulting from the interchange
of the identical antifermions in the final state.  We have checked
the size of the interference term from these contributions and
have found that it is generally small, ranging from about
6\% for a top mass of 160 GeV, to a percent
or less above the $b\thinspace WZ$ threshold.
%

In addition to the full calculation described above, we have
also obtained a ``narrow width'' approximation to the result.
We define this approximation to consist of constraining the $W$ and
$Z$ invariant masses to
their central values by introducing a delta
function into the phase space integral.
The result, when normalized to the appropriate branching ratios,
should be the same at the stable particle result, except for (small)
correction terms proportional to $\Gamma_W/m_W$ or $\Gamma_Z/m_Z$.
Near threshold, this approximation differs from the full result
as the contributions from that portion of phase space where the $W$ and
$Z$ are off mass shell become important.
Below threshold in this approximation, the rate is identically zero.
Note that the cut on the invariant mass of the $e^{+}e^{-}$ pair is
irrelevant in this approximation.
By examining the ratio of the full calculation to the narrow width
result, we get a measure of the size of the enhancement for masses
just above threshold.

In the discussion and plots that follow, we define
\begin{equation}
\Gamma(t\rightarrow b\thinspace W^{+}Z) \equiv
{
{ \Gamma(t\rightarrow b\thinspace\mu^{+}\nu_{\mu} e^{+} e^{-})}
\over
{ {\cal B}(W^{+}\rightarrow\mu^{+}\nu_{\mu})
  {\cal B}(Z\rightarrow  e^{+} e^{-}) }
},
\label{NormalizedRate}
\end{equation}
where
${\cal B}(W^{+}\rightarrow\mu^{+}\nu_{\mu})$
and
${\cal B}(Z\rightarrow  e^{+} e^{-})$
are the experimental
$W$ and $Z$ branching ratios respectively~\cite{caveat}.
We compare this quantity to the
dominant decay mode of a heavy top quark in the Standard Model,
which is $t \rightarrow b\thinspace W^{+}$.
The rate for this decay is given by~\cite{Bigi}
\begin{eqnarray}
\Gamma(t\rightarrow b\thinspace W^{+}) =
{ {G_F}\over{\sqrt{2}} }
{
{m_W^2}
\over
{ 8\pi m_t }
}
\vert V_{tb}\vert^2 &&
\Biggl[
{
{(m_t^2-m_b^2)^2}
\over
{m_t^2m_W^2}
} + {
{m_t^2+m_b^2-2m_W^2}
\over
{m_t^2}
}
\Biggr]
\cr\noalign{\vskip10pt} &&
\negthinspace\negthinspace\negthinspace\times
\sqrt{
\bigl[ m_t^2-(m_W{+}m_b)^2 \bigr] \thinspace
\bigl[ m_t^2-(m_W{-}m_b)^2 \bigr]
}\thinspace ,
\end{eqnarray}
where $G_F$ is the Fermi coupling constant, and $V_{tb}$ is
the CKM matrix element.
The values we have used for the various
masses, widths, coupling constants, and branching ratios
are collected in Table~\ref{Parameters}.

In Figure~\ref{WBZFixedCut}, we have plotted the ratio
$\Gamma(t\rightarrow b\thinspace W^{+}Z)
/\Gamma(t\rightarrow b\thinspace W^{+})$
for a top mass range of 140--260 GeV.
Note that $V_{tb}$ cancels in this ratio.
The solid (dotted) curve
is the result of the full (narrow width) calculation.
The large enhancement of the decay rate near threshold
is clearly evident.  Some representative values for this ratio
appear in Table~\ref{ZResultsTable}.

As has already been stated, the narrow width approximation normalized
as indicated by Eq.~\ref{NormalizedRate} should
reproduce the stable particle result, except
for small corrections.  We find, however, that our result
is in disagreement with the DNP curve \cite{Decker}.  In fact, the
two curves do not even have the same shape.
Our result is  larger by a factor of between 1.6 and 2 over the range
$200 \enspace{\rm GeV} \le m_t \le 300\enspace{\rm GeV}$.
Nevertheless, because we are in agreement with a concurrent,
independent calculation by Ellis~\cite{Ellis}
of $t \rightarrow b\thinspace W^{+} Z$
in the stable particle approximation,
we believe that our curve is indeed correct.

Figure~\ref{WBZVariableCut} explores the effect of varying the
value of the cut on the invariant mass of the $Z$ decay products,
for selected values of the top quark mass.
As would be expected from simple phase space considerations, the
effect of the photon is most visible for low top masses, especially
below the $b\thinspace WZ$ threshold.  In all cases, the choice
$E_{cut} = 0.8\thinspace m_Z$, which corresponds to just over
seven widths below the $Z$ peak, captures the vast majority
of the $Z$ events, while avoiding excessive background
events from the photon.

\section{The Process $\noexpand\lowercase{t} \rightarrow
\noexpand\lowercase{b}\thinspace W H$}

The second decay chain which we wish to consider is
\begin{equation}
 t \rightarrow b\thinspace W^{+}\thinspace H \thinspace ;
\quad W^{+} \rightarrow   \mu^{+} \nu_{\mu}.
\end{equation}
Since the width of the Standard Model Higgs is only about
5 MeV for masses below the $W$-pair threshold, we do not include
its width in our calculation.  This is a much simpler process
than~(\ref{tbWZ}), as there are just four Feynman diagrams
in the Standard Model:  the Higgs may be attached
to the $t$, $b$, $W$, or $\mu^{+}$.  Since we work in the approximation
$m_{\mu} = 0$, only the first three of these
diagrams actually contribute.  Although small, we retain the
contribution from the $bbH$ vertex.

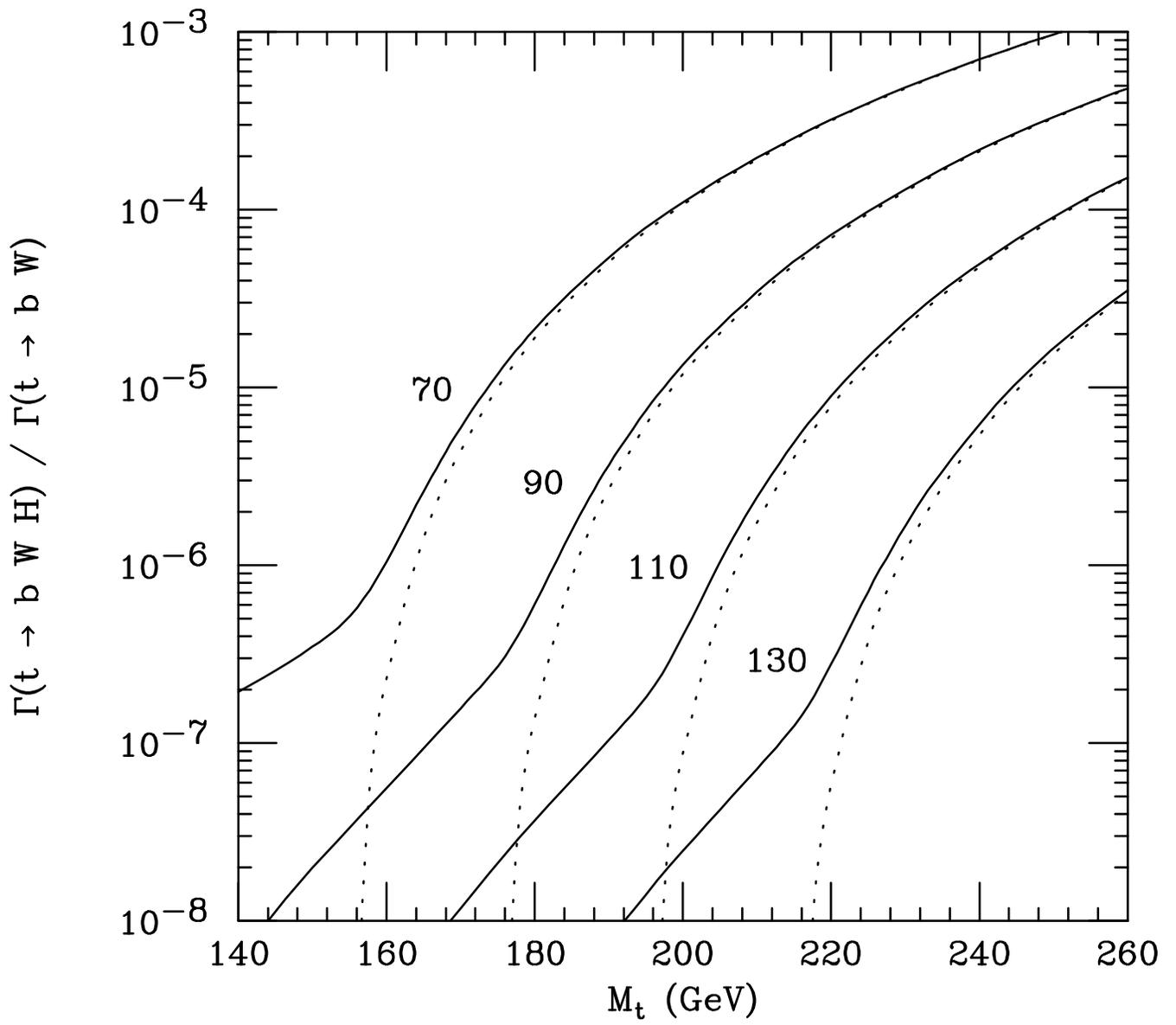
\begin{figure}
\caption[]{The ratio
$\Gamma(t\rightarrow b\thinspace W^{+}H)
/\Gamma(t\rightarrow b\thinspace W^{+})$
versus $m_t$,
for Higgs masses of 70, 90, 110, and 130 GeV.  In each case,
the solid line is the full calculation including the $W$
width effects, while the dotted line is the
narrow width approximation.  }

\label{HiggsPlot}
\end{figure}

Defining
\begin{equation}
\Gamma(t\rightarrow b\thinspace W^{+}H) \equiv
{
{\Gamma(t\rightarrow b\mu^{+}\nu_{\mu} H)}
\over
{{\cal B}(W^{+}\rightarrow \mu^{+} \nu_{\mu})}
}
\end{equation}
and proceeding as before, we plot as Fig.~\ref{HiggsPlot} the ratio
$\Gamma(t\rightarrow b\thinspace W^{+}H)
/\Gamma(t\rightarrow b\thinspace W^{+})$
in both the full and narrow width
approximation,
as a function of $m_t$
for selected Higgs masses. 
The narrow width curves plotted here do indeed agree with the
literature~\cite{Decker,DeckerH}.  Some selected values
for this ratio are listed in Table~\ref{HiggsTable}.

Although the rates are somewhat more favorable in this case than
for $b\thinspace WZ$, provided
$m_H < m_Z$, the difficulty in actually
seeing the Higgs associated
with this decay makes observation of this mode in the near future
unlikely.


\section{Conclusions}

We have performed a calculation of the decays
$t \rightarrow b\thinspace W^{+}Z$ and
$t \rightarrow b\thinspace W^{+}H$ including the subsequent
decays of the gauge bosons.  For the $b\thinspace WZ$ final
state, we disagree with the existing stable particle
calculation~\cite{Decker} well above threshold.  We agree with the
previously published results~\cite{Decker,DeckerH} for the
$b\thinspace WH$ final state in this limit.  In both decays, we find a
significant enhancement in the rate for interesting values
of the top quark mass.  This enhancement is not sufficient, however,
to render either decay visible to the current generation
of experiments.  Thus, we conclude that the observation of
either decay would signal the existence of physics beyond the
Standard Model.

\acknowledgements

We would like to thank Keith Ellis for useful discussions.
This work was performed at the Fermi National Accelerator
Laboratory, which is operated by Universities Research Association,
Inc., under contract DE-AC02-76CHO3000 with the U.S. Department
of Energy.


\begin{figure}[h]

\vskip0.2cm

\caption[]{The ratio
$\Gamma(t\rightarrow b\thinspace W^{+}Z)/
  \Gamma(t\rightarrow b\thinspace W^{+})$
versus $m_t$,
with an $e^{+}e^{-}$ invariant mass cut of $0.8\thinspace m_Z$.
The solid line is the full calculation including the $W$ and $Z$
width effects, while the dotted line is the
narrow width approximation.  }
\label{WBZFixedCut}
\end{figure}

\begin{figure}
\caption[]{The ratio
$\Gamma(t\rightarrow b\thinspace W^{+}Z)
/\Gamma(t\rightarrow b\thinspace W^{+})$
versus the $e^{+}e^{-}$ invariant mass cut described in the text,
for top quark masses of 160, 176, 200, and 250 GeV.}
\label{WBZVariableCut}
\end{figure}

\begin{table}
\caption{Input values used in calculation of
$t \rightarrow b\thinspace W^{+}Z(H)$.
\label{Parameters}}
\begin{tabular}{cccccccc}
&&&$m_W$ &  $80\enspace{\rm GeV}$&&& \\
&&&$m_Z$ &  $91\enspace{\rm GeV}$&&& \\
&&&$m_b$ &  $\phantom{0}5\enspace{\rm GeV}$&&& \\
&&&$\Gamma_W$ &  $2.11\enspace{\rm GeV}$&&& \\
&&&$\Gamma_Z$ &  $2.49\enspace{\rm GeV}$&&& \\
&&&$G_F$ &  $1.166 \times 10^{-5}\enspace{\rm GeV}^{-2}$&&& \\
&&&$\sin^2\theta_W$ &  $0.2319$&&& \\
&&&${\cal B}(W^{+}\rightarrow\mu^{+}\nu_{\mu})$ & $0.107$&&& \\
&&&${\cal B}(Z\rightarrow  e^{+} e^{-})$ & $0.03367$&&& \\
\end{tabular}
\end{table}

\begin{table}
\caption{The ratio
$\Gamma(t\rightarrow b\thinspace W^{+}Z)/
 \Gamma(t\rightarrow b\thinspace W^{+})$ for the
full calculation (F),  the narrow
width approximation (NW), and their ratio (F/NW).
\label{ZResultsTable}}
\begin{tabular}{cddd}
$m_t$ (GeV) & F & NW & F/NW \\
\tableline
150     & 6.68$\times 10^{-8}$ &   --                     & --  \\
160     & 1.60$\times 10^{-7}$ &   --                     & --  \\
177     & 7.47$\times 10^{-7}$ & 1.45$\times 10^{-9}$     & 516. \\
185     & 2.14$\times 10^{-6}$ & 5.57$\times 10^{-7}$     & 3.84 \\
200     & 1.36$\times 10^{-5}$ & 1.00$\times 10^{-5}$     & 1.35 \\
250     & 2.56$\times 10^{-4}$ & 2.54$\times 10^{-4}$     & 1.01 \\
\end{tabular}
\end{table}

\begin{table}
\caption{The ratio
$\Gamma(t\rightarrow b\thinspace W^{+}H)/
 \Gamma(t\rightarrow b\thinspace W^{+})$ with a 90 GeV Higgs for the
full calculation (F),  the narrow
width approximation (NW), and their ratio (F/NW).
\label{HiggsTable}}
\begin{tabular}{cddd}
     $m_t$ (GeV) & F & NW &  F/NW \\
\tableline
      160 & 5.75$\times 10^{-8}$ &   --                   & --   \\
      176 & 3.06$\times 10^{-7}$ & 2.70$\times 10^{-9}$   & 113. \\
      185 & 1.57$\times 10^{-6}$ & 8.67$\times 10^{-7}$   & 1.81 \\
      200 & 1.34$\times 10^{-5}$ & 1.19$\times 10^{-5}$   & 1.13 \\
      250 & 3.32$\times 10^{-4}$ & 3.28$\times 10^{-4}$   & 1.01 \\
\end{tabular}
\end{table}

\begin{references}

\bibitem[*]{GDMemail} Electronic address:  gdm@fnth03.fnal.gov
\bibitem[\dagger]{SJPemail} Electronic address:  parke@fnalv.fnal.gov

\bibitem{CDFtop} F.~Abe, {\it et. al.},
Phys. Rev. {\bf D50}, 2966 (1994);
Phys. Rev. Lett. {\bf 73}, 225 (1994).

\bibitem{LEPfit}  Aguilar-Benitez, {\it et. al.}, Phys. Rev.
{\bf D50}, 1173 (1994).

\bibitem{Decker}  R. Decker, M. Nowakowski, and A. Pilaftsis,
Z. Phys. {\bf C57}, 339 (1993).

\bibitem{DeckerH}  R. Decker, M. Nowakowski, and A. Pilaftsis,
Mod. Phys. Lett. {\bf A6}, 3491 (1991); Erratum {\bf A7}, 819 (1992).

\bibitem{caveat} Strictly speaking, we should use an appropriately
weighted sum composed of results from all possible decay modes of
the $W$ and $Z$.  Provided that we choose the $e^{+}e^{-}$
invariant mass cut so as to minimize the contamination from
the photon, our definition is accurate
up to errors of $O(\alpha_s)$.  We have verified that the results
obtained by substituting other decay modes for the $W$ or $Z$
are indeed consistent at this level.

\bibitem{Bigi} I. Bigi, Y. Dokshitzer, V. Khoze, J. K\"uhn,
and P. Zerwas, Phys. Lett. {\bf 181B}, 157 (1986).

\bibitem{Ellis}  R.K. Ellis, private communication.

\end{references}
\end{document}